\documentclass[a4paper,12pt]{article}

\usepackage{graphics, epsfig}

\renewcommand{\theequation}{\thesection.\arabic{equation}}
\begin{document}


\newcommand{\ba} {\begin{eqnarray}}
\newcommand{\ea} {\end{eqnarray}}
\newcommand{\be}{\begin{equation}}
\newcommand{\ee}{\end{equation}}
\newcommand{\n}[1]{\label{#1}}
\newcommand{\eq}[1]{eq.(\ref{#1})}
\newcommand{\ind}[1]{\mbox{\tiny{#1}}}
\renewcommand\theequation{\thesection.\arabic{equation}}

\newcommand{\nn}{\nonumber \\ \nonumber \\}
\newcommand{\nl}{\\  \nonumber \\}
\newcommand{\pr}{\partial}
\renewcommand{\vec}[1]{\mbox{\boldmath$#1$}}

\newcommand{\ben}{$$}
\newcommand{\een}{$$}
\author{C. Barrab\`es\thanks{E-mail : barrabes@celfi.phys.univ-tours.fr}\\
\small CNRS/UMR 6083, Universit\'e F. Rabelais, 37200 TOURS, France\\
P. A. Hogan\thanks{E-mail : peter.hogan@ucd.ie}\\ \small
Math. Phys. Dept., University College Dublin, Belfield, Dublin 4, Ireland}

\title{Singular Hypersurfaces in Einstein-Gauss-Bonnet Theory of Gravitation}
\date{}
\maketitle

\begin{abstract}
We present a general formalism for describing singular
hypersurfaces in the Einstein theory of gravitation with a
Gauss-Bonnet term. The junction conditions are given in a form
which is valid for the most general embedding and matter content
and for coordinates chosen independently on each side of the hypersurface.
The theory is applied to both a time-like and a light-like hypersurface
in brane-cosmology.
\end{abstract}
\vskip 2truepc\noindent PACS number(s): 04.20.Cv
\thispagestyle{empty}
\newpage

\setcounter{equation}0

Space--times with noncompact extra dimensions are presently
extensively studied, in particular in the context of
brane-cosmology. These models, which are inspired from high-energy
physics, string theory and $M$-theory, lead to a brane-world
picture. In the most popular model \cite{RS}-\cite{BDL} our
ordinary four-dimensional space--time is the history of a
three-dimensional brane in a five-dimensional space--time (the
bulk). All matter and gauge fields, except gravity,
 are confined to a 3-brane whose history is a time--like shell
embedded in a 5-dimensional anti-de Sitter space-time. An
additional feature of string theory is that the low-energy
effective action contains terms which are quadratic in the
curvature. The Gauss-Bonnet term, $L_{GB}=
R_{\alpha\beta\rho\sigma}\, R^{\alpha\beta\rho\sigma} -4
R_{\alpha\beta}\,R^{\alpha\beta} + R^2$, has then been included in
the Einstein-Hilbert action.

When the Gauss-Bonnet term is introduced into the brane-world
models the field equations of the $5$-D space-time are modified
and one also expects the field equations on the brane to be
modified. Several authors have studied this problem \cite{ND} and
have for the most part only considered the particular situation
corresponding to brane-cosmology ($Z_2$-symmetry for the
embedding, anti-de Sitter geometry for the bulk and perfect fluid
on the brane). Their results concerning the modifications of the
Friedmann equation on the brane are still controversial.
Consequently it is important to have a systematic treatment of the
junction conditions involved. The approach taken by some of the authors in
\cite{ND} uses a Lagrangian with appropriate boundary terms (the
Lagrangian method for shells in General Relativity 
is not free of ambiguities-- see for example \cite{Hay}) whereas
our approach starts with the field equations and makes use of the
Gauss--Codazzi equations. \emph{It is the purpose of this paper to
present a general formalism for the description of the
junction-conditions on a singular hypersurface when the
Gauss-Bonnet term is present}. Our formalism places no
restrictions on the matter content or the geometry of the outer
space or on the matter content of the shell. The only restriction
is the basic embedding condition that the induced metrics on the
hypersurface coincide. This is achieved by extending to the
Einstein-Gauss-Bonnet theory a previous work \cite{BI} where the
junction-conditions in the Einstein theory were given for an
arbitrary hypersurface (time--like, space--like or light--like).
Our general formalism shows that, in spite of the presence of
quadratic terms in the curvature, no regularisation of the Dirac
$\delta$-function is required and a well--defined set of junction
conditions is obtained. We then apply our results to brane-world
cosmology and consider both a time--like and a null hypersurface.

Our results, summarized in (\ref{2.17})--(\ref{2.21}) below, are
completely general and do not assume any symmetries (of the
space--time or the embedding) in contradistinction to the works
cited in \cite{ND} where, in particular $Z_2$-symmetry is
introduced at the outset. This is a source of confusion with
regard to the distinction between \emph{the average} and \emph{the
jump} of a quantity which is discontinuous across the hypersurface
(equations such as (\ref{2.13}) and (\ref{2.19}) below contain
both averages and jumps of quantities). This appears to be the
fundamental origin of the discrepancy between our results and
those cited in \cite{ND}.

We consider a $5$-dimensional space--time $\cal M$ with a system
of local coordinates $\{x^{\mu}\}$, $\mu =0,1,2,3,4$. The
components of tensors on $\cal M$ in this coordinate system will
be identified by an index $5$. For example the metric tensor
components will be denoted ${}^5 g_{\mu\nu}$. The field equations
are
\be
{}^5 G_{\mu\nu} + \Lambda_5\,\, {}^5 g_{\mu\nu} + 2\alpha
H_{\mu\nu} =\kappa_5\,\, {}^5 T_{\mu\nu}\ , \n{1.1}
\ee
where $^5 G_{\mu\nu}$ is the Einstein tensor calculated with the metric
tensor ${}^5g_{\mu\nu}$ , $\Lambda_5$ is the cosmological
constant, $\alpha$ is a coupling constant and $H_{\mu\nu}$ is the
Lovelock tensor which is given by
\be
\begin{array}{l}
H_{\mu\nu}= {}^5 R\,\, {}^5 R_{\mu\nu}
-2\,\,{}^5 R_{\mu}{}^{\lambda}\,\, {}^5 R_{\lambda\nu}
-2\,\,{}^5 R^{\alpha\beta}\,\, {}^5 R_{\alpha\mu\beta\nu}
+ {}^5 R_{\mu\rho\kappa\lambda}\,\,{}^5 R_{\nu}{}^{\, \rho\kappa\lambda}\\
\displaystyle -\frac{g_{\mu\nu}}{4}
({}^5 R_{\alpha\beta\rho\sigma}\,\,{}^5 R^{\alpha\beta\rho\sigma}
-4\,\, {}^5 R_{\alpha\beta}\,\, ^5 R^{\alpha\beta} +{}^5 R^2)\,. \,\n{1.2}
\end{array}
\ee
Here ${}^5R_{\mu\rho\kappa\lambda}\,,\ {}^5R_{\mu\nu}\,,\ {}^5R$
are the components of the Riemann tensor, Ricci tensor and Ricci
scalar respectively calculated with the metric tensor
${}^5g_{\mu\nu}$. In the right hand side of the field equations
the coefficient $\kappa_5$ is the $5$-D gravitational constant and
$^5 T_{\mu\nu}$ is the stress-energy tensor describing the matter
content (of the bulk and the brane in the brane-world language).

The space--time manifold $\cal M$ is divided into two domains
${\cal M}^{\pm}$ by a singular hypersurface $\cal N$
on which the metric tensor is only $C^0$. As a
consequence of this the Riemann curvature tensor contains a Dirac
$\delta$-term with support on $\cal N$.
Each domain ${\cal M}^{\pm}$ admits a metric tensor $^5 g^{\pm}$  and
 all quantities referring to ${\cal M}^{\pm}$
will be denoted by an index ${\pm}$.
We denote the jumps across $\cal N$ of a quantity
$F^{\pm}$ defined on ${\cal M}^{\pm}$ by $[F]= F^+\vert _{\cal N}
\,-\,F^-\,\vert _{\cal N}$, where $\vert _{\cal N}$ indicates that
$F^{\pm}$ is to be evaluated on the $\pm$ sides of ${\cal N}$
respectively.

\section{Junction conditions on a timelike or spacelike hypersurface}
\setcounter{equation}0

Let $\{ x^{\mu}\}$ be a local coordinate system covering both
sides of the hypersuface $\cal N$ in terms of which the
components of the metric tensor are continuous across $\cal N$.
Let $\Phi (x)=0$ be the equation of $\cal N$ in these
coordinates, with
$\Phi >0(<0)$ in ${\cal M}^+$(${\cal M}^-$). Greek
indices take values $0,1,2,3,4$ and the components of the metric
tensor are $g^{\pm}_{\mu\nu}$ in ${\cal M}^{\pm}$ respectively. If
$F^{\pm}(x)$ are two quantities (the components of a tensor for
example) defined on ${\cal M}^{\pm}$ respectively, we define the
hybrid quantity $\tilde{F}$ by
\be
\tilde{F}(x)= F^+\,\Theta (\Phi)+F^-\,(1-\Theta (\Phi))\  ,
\n{2.1}
\ee
where $\Theta (\Phi )$ is the Heaviside step function
which is equal to unity (zero) when $\Phi >0 (<0)$.
Thus in particular for the metric tensor we have $^5
\tilde{g}_{\mu\nu}$ defined and since the metric is continuous
across $\cal N$ we can write $[^5 g_{\mu\nu}]=0$. As a result of
the definition (\ref{2.1}) we have for two quantities $F^{\pm}$
and $G^{\pm}$ the product rule, $\tilde{F} \tilde{G}=\tilde{FG}$,
because $\Theta (\Phi)(1-\Theta (\Phi))$ vanishes
distributionally. The normal to the hypersurface has components
\be
n_{\mu}= \chi^{-1}(x) \,\partial_{\mu} \Phi (x)\, ,\n{2.2}
\ee
where $\chi (x)$ is a normalizing factor such that
\be
n\cdot n={}^5g^{\mu\nu} n_{\mu}n_{\nu}\,|_{\pm}=\epsilon\,
,\n{2.3}
\ee
with $\epsilon=+1\,(-1)$ if the hypersurface is
time--like (space--like). With these definitions the partial
derivative of $\tilde{F}$ takes the form
\be
\partial_{\mu} \tilde{F} = \tilde{\partial_{\mu}F}\,+\,
[F]\,\chi n_{\mu}\,\delta (\Phi)\, .\n{2.4}
\ee
A singular term
proportional to the Dirac $\delta$-funtion appears whenever $F$ is
discontinuous across the hypersurface $\cal N$. The metric and its
tangential derivatives are continuous across $\cal N$ but its
transverse derivatives are not. To characterize the
discontinuities in the transverse derivatives of the metric tensor
we define the symmetric tensor $\gamma_{\mu\nu}$ by
\be
[\partial_{\alpha}\,{}^5g_{\mu\nu}]=\epsilon\,n_{\alpha}\,\gamma_{\mu\nu}\,
.\n{2.5}
\ee
The tensor $\gamma_{\mu\nu}$ is only defined
\emph{on} $\cal N$ and its projection onto $\cal N$
is unique. Thus $\gamma_{\mu\nu}$ is free up to the gauge
transformation, $\gamma_{\mu\nu}\rightarrow\gamma _{\mu\nu}'=\gamma_{\mu\nu} +
v_{\mu}n_{\nu}+ n_{\mu}v_{\nu}$, where $v$ is an
arbitrary vector field on $\cal N$.
This gauge freedom can
always be used to have $\gamma_{\mu\nu} n^{\nu}=0$
whenever $\cal N$ is not lightlike. Using
(\ref{2.5}) we find that the Christoffel symbols satisfy
${}^5\Gamma^{\lambda}_{\mu\nu}={}^5
\tilde{\Gamma}^{\lambda}_{\mu\nu}$ and
$[{}^5\Gamma^{\lambda}_{\mu\nu}]=\epsilon\,\gamma^{\lambda}_{(\mu}n_{\nu)}
-\frac{\epsilon}{2}\gamma_{\mu\nu}n^{\lambda}$,
with round brackets around indices denoting symmetrisation. Then using
(\ref{2.4}) the Riemann curvature tensor
${}^5R_{\kappa\lambda\mu\nu}$ can be decomposed into the sum of a
tilde-term defined as in (\ref{2.1}) and a term containing a Dirac
$\delta$-function:
\be
{}^5R_{\kappa\lambda\mu\nu}={}^5\tilde{R}_{\kappa\lambda\mu\nu}+
\hat{R}_{\kappa\lambda\mu\nu}\,\epsilon \chi\delta(\Phi)\ ,
\n{2.8}
\ee
where
\be
\hat{R}_{\kappa\lambda\mu\nu}=2\,n_{[\kappa} \gamma_{\lambda][\mu}
n_{\nu]}\ .\n{2.9}
\ee
The square brackets here around indices
denote skew--symmetrisation. A similar decomposition therefore
exists for the Ricci tensor ${}^5R_{\mu\nu}$, the Ricci scalar
${}^5R$ and the Einstein tensor
${}^5G_{\mu\nu}={}^5R_{\mu\nu}-\frac{1}{2}\,{}^5R\,{}^5
g_{\mu\nu}$. The singular part of the Einstein tensor is given by
\be
\hat{G}_{\mu\nu}=\gamma_{(\mu}n_{\nu)}-\frac{\gamma}{2}\,n_{\mu}\,n_{\nu}
-\frac{\gamma^\dagger}{2}\,\,{}^5g_{\mu\nu}
-\frac{\epsilon}{2}\,(\gamma_{\mu\nu}-\gamma\,\,{}^5g_{\mu\nu})\ ,
\n{2.10}
\ee
where we have introduced
$\gamma\equiv \gamma^{\mu}_{\mu}$,
$\gamma_{\mu}\equiv\gamma_{\mu\nu}\,n^{\nu}$ and
$\gamma^\dagger\equiv\gamma_{\mu}\,n^{\mu}$
(recall that one can always choose the gauge such
that $\gamma_{\mu}=\gamma^{\dagger}=0$).

If we now use (\ref{2.8}) to calculate $H_{\mu\nu}$ in (\ref{1.2})
undefined $\delta^2$-terms will appear since $H_{\mu\nu}$ is
quadratic in the Riemann tensor. However one can show that 
these terms simply disappear. To see this we first note that for
any tensor $A$ having the form $A=\tilde
{A}+\hat{A}\,\chi\,\epsilon\,\delta (\Phi)$ and any other tensor
$B$ having the same form their product $AB$ contains a
$\delta^2$-term with coefficient $\hat{A}\hat{B}\chi^2
\epsilon^2$. If this is applied to the calculation of $H_{\mu\nu}$
and use is made of the expressions for $\hat{R},\
\hat{R}_{\mu\nu}$ and $\hat{R}_{\kappa\lambda\mu\nu}$ given or
derived from (\ref{2.9}) we find that the total contribution of
all products of the type $\hat{A}\hat{B}$ is zero. Therefore the
coefficient of the $\delta^2$ term in $H_{\mu\nu}$ vanishes and we
can write
\be
H_{\mu\nu}=\tilde{H}_{\mu\nu}+\hat{H}_{\mu\nu}\,\epsilon
\chi\,\delta (\Phi)\ ,\n{2.12}
\ee
where $\hat{H}_{\mu\nu}$ is given by
\be
\label{2.13}
\begin{array}{l}
\displaystyle \hat{H}_{\mu\nu}=\hat{R}_{\mu\nu}\,\,{}^5\bar{R}
+\hat{R}\,\,{}^5\bar{R}_{\mu\nu}
-2\,(\hat{R}^{\alpha\beta}\,\,{}^5\bar{R}_{\alpha\mu\beta\nu}+
\hat{R}_{\alpha\mu\beta\nu}\,\,{}^5\bar{R}^{\alpha\beta})\\
\\
\displaystyle +\hat{R}_{\mu\kappa\rho\lambda}\,\,
{}^5\bar{R}_{\nu}{}^{\kappa\rho\lambda}+
\hat{R}_{\nu}{}^{\kappa\rho\lambda}\,\,
{}^5\bar{R}_{\mu\kappa\rho\lambda}-2\,(\hat{R}_{\mu}{}^{\lambda}\,\,
{}^5\bar{R}_{\nu\lambda}+\hat{R}_{\nu}{}^{\lambda}\,\,
{}^5\bar{R}_{\mu\lambda})\\
\\
\displaystyle -\frac{{}^5g_{\mu\nu}}{2}(
\hat{R}_{\alpha\beta\rho\sigma}\,\,{}^5
\bar{R}^{\alpha\beta\rho\sigma}-4\,\hat{R}_{\alpha\beta}\,\,{}^5
\bar{R}^{\alpha\beta}+\hat{R}\,\,{}^5\bar{R})\ .
\end{array}
\ee
In those expressions  $\tilde{H}_{\mu\nu}$ has the general form
(\ref{2.1}) and the bar denotes the `average' of a quantity which is
discontinuous across $\cal N$ (thus $\bar{A}= (A^+\vert _{\cal N}
+A^-\vert _{\cal N})/2$ for any $A$ for which $[A]\neq 0$). Also the
property $\Theta (x)\,\delta (x)=\frac{1}{2}\,\delta (x)$,
which is distributionally valid, has been
used. When these results are substituted into the field equations
(\ref{1.1}) we find that the stress-energy tensor on the right
hand side of the equations is decomposed into a tilde term and a
singular term with the latter indicating the presence, in general,
of a thin shell with history $\cal{N}$. Thus
\be
^5 T_{\mu\nu} =  {}^5 \tilde {T}_{\mu\nu} + S_{\mu\nu} \chi\delta
(\Phi)\, ,\n{2.14}
\ee
where the tensor $S_{\mu\nu}$ is the surface
stress-energy tensor of the shell. Identifying the singular terms
on each side of the field equations we have
\be
\kappa_5 \, S_{\mu\nu}=\,\epsilon\, \hat{G}_{\mu\nu} +
2\,\alpha \, \epsilon \,\hat{H}_{\mu\nu}\ ,\n{2.15}
\ee
where $\hat{G}_{\mu\nu}$ is given by (\ref{2.10})
and $\hat{H}_{\mu\nu}$ by (\ref{2.12}).
Since (\ref{2.10}) and (\ref{2.13}) lead to
$\hat{G}_{\mu\nu}\,n^{\nu}=0$ and $\hat{H}_{\mu\nu}\,n^{\nu}=0$
we also have $S_{\mu\nu}\,n^{\nu}=0$ .
This tangential property for  $S_{\mu\nu}$, and the vanishing of the
$\delta^2$-terms in $H_{\mu\nu}$ are both a direct consequence of the
particular form of the Gauss-Bonnet term.

The Eq.(\ref{2.15}) represents the junction conditions on $\cal N$
in a system of coordinates covering both sides of the
hypersurface. This presentation is equivalent to the usual
Israel junction conditions formalism \cite{WI} as there is a
direct relation between the tensor $\gamma_{\mu\nu}$ and the jump
in the extrinsic curvature of $\cal N$. The formalism based on the extrinsic
curvature has the advantage of allowing the 5-D space--time
coordinates to be chosen freely and independently on each side of
the hypersurface. Let $\{x_{\pm}^{\mu}\}$ be local
chartss for the domains $\cal{M}^{\pm}$ and introduce on the
hypersurface $\cal{N}$ four intrinsic coordinates $\{\xi^a\}$ with
$a=0,1,2,3$. The four holonomic tangent basis vectors
$e_{(a)}=\partial /\xi^a$ have components
$e^{\mu}_{(a)}\vert_{\pm}= \partial x^{\mu}_{\pm}/\partial \xi^a$,
and the induced metrics $g_{ab}={}^5 g_{\mu\nu} e^{\mu}_{(a)}
e^{\nu}_{(b)}\vert_{\pm}$ match on $\cal N$.
The extrinsic curvature is defined by
$K_{ab}=-n_{\mu}\,e^{\mu}_{(a)|\lambda}\,e^{\lambda}_{(b)}$ and
takes different values $K^{\pm}_{ab}$ on each side of $\cal{N}$.
Here the stroke denotes covariant differentiation
associated with the five dimensional
metric tensor on either side of $\cal N$. It can be shown that
$\gamma_{ab} \equiv 2[K_{ab}]$ is the projection on $\cal N$ of
the tensor $\gamma_{\mu\nu}$ introduced in (\ref{2.5}), thus
$\gamma_{ab}=\gamma_{\mu\nu}\,e^{\mu}_{(a)}\,e^{\nu}_{(b)} $.
The projection of the surface
stress--energy tensor onto $\cal N$ is given by
$S_{ab}=S_{\mu\nu}\,e^{\mu}_{(a)}\,e^{\nu}_{(b)}$. In similar
fashion the projections $\hat G_{ab}$ and $\hat H_{ab}$ onto $\cal
N$ of the singular parts of $G_{\mu\nu}$ and $H_{\mu\nu}$ are
defined. The junction conditions (\ref{2.15}) now read
\be
\kappa_5\, S_{ab}=\epsilon \hat{G}_{ab}+
2\,\alpha\,\epsilon \hat{H}_{ab}\
.\n{2.17}
\ee
Using the Gauss-Codazzi equations and their
contracted forms in (\ref{2.10}) and (\ref{2.13}) we find that
\be
\epsilon\,\hat{G}_{ab}=-[K_{ab}]+[K]\,g_{ab}\ ,\n{2.18}
\ee
\be
\epsilon\, \hat{H}_{ab}=-2\,[K^{cd}]\,({}^*R^*_{acbd}
-{}^*\bar{r}^*_{acbd})\ .\n{2.19}
\ee
In the last relation ${}^*
R^*_{acbd}$ is the left and right dual of the intrinsic Riemann
tensor of the hypersurface $\cal{N}$ and is given by
\be
{}^*R^*_{acbd}=-R_{acbd}+2\,g_{a[b}R_{d]c}-2\,g_{c[b}R_{d]a}-R\,
g_{a[b}g_{d]c}\ .\n{2.20}
\ee
The final term in (\ref{2.19}) is
the left and right dual of the average on $\cal{N}$ of
\be
r_{acbd}= K_{ab} K_{cd} -K_{ad} K_{bc}\ ,\n{2.21} \ee which has
the same algebraic symmetries as $R_{acbd}$. When the Gauss-Bonnet
term is absent ($\alpha =0$) only $\hat{G}_{ab}$, which simply
depends on $[K_{ab}]$, appears in the junction relations and we
recover the usual Israel conditions \cite{WI}. This limit $\alpha
=0$ is not in, for example, the cited
work of Davis \cite{ND}, if this work is interpreted using the
standard convention for the normal to be found in  \cite{BI} or \cite{WI}.  
The Gauss--Bonnet contribution to the
junction relations is contained in $\hat{H}_{ab}$ which depends
not only on $[K_{ab}]$, but also on the intrinsic curvature tensor
$R_{acbd}$ of $\cal{N}$ and on the average of products of the
extrinsic curvature. The equation (\ref{2.17}) describes the
evolution of the shell once a choice is made of the surface
stress-energy tensor $S_{ab}$. In brane-cosmology it provides the
new field equations on the brane. Two other relations can be
derived by considering the jump of the field equations
(\ref{1.1})-(\ref{1.2}) across $\cal N$ contracted with
$n^{\mu}\,e^{\nu}_{(a)}$ or $n^{\mu}\,n^{\nu}$, and making use of
the Gauss-Codazzi equations:
\be
S^b{}_{a;b} =-[{}^5T_{\mu\nu}\,n^{\mu}\,e^{\nu}_{(a)}]+
\frac{2\,\alpha}{\kappa_5}
\left (\epsilon \hat{H}^b{}_{a;b} +[H_{\mu\nu}\,n^{\mu}\,e^{\nu}_{(a)}]
\right )\ ,\n{2.22}
\ee
\be
S_{ab}\,\bar{K}^{ab}=[{}^5T_{\mu\nu}\,n^{\mu}\,n^{\nu}]-
\frac{\epsilon}{\kappa_5}[\Lambda_5]
+\frac{2\,\alpha}{\kappa_5}
\left (\epsilon \hat{H}_{ab}\,\bar{K}^{ab}  -[H_{\mu\nu}\,n^{\mu}\,n^{\nu}]
\right )\ .\n{2.23}
\ee
where the semicolon here denotes the covariant derivative
associated with $g_{ab}$. For a time--like shell the $a=0$
component of (\ref{2.22}) gives the energy conservation equation.

As an illustration of our general formalism we consider
brane-cosmology and the most popular case where the $5$-D
space--times have anti-de Sitter geometry, the embedding is $Z_2$
symmetric and the matter on the brane is a perfect fluid with
proper--density $\rho$ and pressure $p$. In this case
$\epsilon =1$ as $\cal{N}$ is taken to be
time--like. The assumption of $Z_2$ symmetry simplifies the
expression for the average of the tensor $r_{acbd}$ defined above
to read
\be
\bar{r}_{acbd}=\frac{1}{4}\,\left (
[K_{ab}]\,[K_{cd}]-[K_{ad}]\,[K_{bc}]\right ) \ .\n{2.24}
\ee
The 5-dimensional metric is given via the line--element
\be
ds^2 =-h(r)dt^2 + h(r)^{-1} dr^2 + r^2 d\chi^2 + r^2 f^2_k (\chi)
(d\theta^2 +\sin^2\theta d\phi^2)\ .\n{2.25}
\ee
Here $f_k (\chi)=1,\ \sin\chi ,\ \sinh\chi$ for $k=0,\ 1,\ -1$,
and $h(r)$ is given by \cite{BD}
\be
h(r)=k+\frac{r^2}{4\alpha}\,( 1- \sqrt{A(r)})\ ,\n{2.26}
\ee
where $A(r)=1+\frac{4}{3}\,\alpha\,\Lambda_5 +8\,\alpha\,\frac{m}{r^4}$, with
$m$ being a mass parameter.
The shell is radially moving with the law
of motion $r= a(\tau),\ t=t(\tau )$,
and $\tau$ given by $dt/d\tau =(h+\dot a^2)^{1/2}/h$.
The two functions
$f_k(\chi)$ and $h(r)$ are the same on both sides of $\cal N$ and
the metric on $\cal{N}$ has
the Robertson--Walker form
\be
ds^2\vert_{\cal{N}}=-d\tau^2+a^2(\tau)\,d\chi^2+a^2(\tau)\,f^2_k
(\chi)(d\theta ^2+\sin ^2\theta\,d\phi ^2)\ .\n{2.27}
\ee
Now $S^{ab}$ has the perfect fluid form
$S^{ab}= (\rho + P)\,u^a\,u^b+P\,g^{ab}$ with $u^a=(1,0,0,0)$ the
four-velocity of the fluid. The assumption of $Z_2$ symmetry
implies for the extrinsic curvature $K^+_{ab}=-K^-_{ab}\equiv
K_{ab}$, and because the metric given by (\ref{2.25}) is
spherically symmetric we have
$K^{\chi}_{\chi}=K^{\theta}_{\theta}=K^{\phi}_{\phi}=\zeta\,\sqrt{\dot{a}^2
+ h(a)}/ a$, where $\dot{}=d/d\tau$ and $\zeta ={\rm
sign}\,(n^{\mu}\,\partial_{\mu}r)$. Introducing these properties
into the equation (\ref{2.17}) contracted twice with $u^a$ we
arrive at
\be
\kappa_5\,\rho =-\frac{6
\zeta}{a}\,\sqrt{\{\dot{a}^2+h(a)\}\,A(a)}\ .\n{2.28}
\ee
Squaring
this we obtain the Friedmann equation on the brane when a
Gauss-Bonnet term is present
\be
{\cal H}^2=-\frac{k}{a^2}+\frac{\kappa_5^2\, \rho^2}{36
A(a)}-\frac{1}{4\alpha}\,\{1-\sqrt{A(a)}\}\ .\n{2.29}
\ee
Here
${\cal H}=\dot{a} /a$ is the Hubble parameter. It is easy to check that
in the limit $\alpha\rightarrow 0$ this equation reduces to the
usual Friedmann equation on the brane without the Gauss-Bonnet
term (see for example \cite{BDL}).

\section{Junction conditions for a light--like hypersurface}
\setcounter{equation}0

The general formalism that we have developed for a time--like or a
space--like shell can be adapted to the case of a null shell. We
briefly present the main results here and refer the reader to
\cite{BI} for more details concerning the properties of a
light--like shell. As an illustration we describe a null brane
propagating in the brane-cosmological model already considered in
the previous section.

The hypersurface $\cal N$ is now light--like and therefore its
normal $n$ satisfies (\ref{2.3}), with $\epsilon =0$, and
is tangent to $\cal N$.
The definition (\ref{2.2}) for $n$ still applies but
the function $\chi (x)$ is now arbitrary.
We introduce a
transversal vector $N$ on $\cal N$ such that
$N\cdot n=\eta ^{-1}\neq 0$,
and describe the
discontinuity across $\cal{N}$ in the transverse first derivative
of the metric by the tensor $\gamma_{\mu\nu}$ which is defined by
the same equation (\ref{2.5}) with $\epsilon$ replaced by $\eta$.
The decomposition (\ref{2.8}) of the Riemann tensor still applies with
again $\epsilon$ replaced by $\eta$ and its singular part has the
same expression (\ref{2.9}). However the singular part of the
Einstein tensor (\ref{2.10}) now specializes to only 
the first three terms in (\ref{2.10}). 
Note that here the gauge freedom for $\gamma_{\mu\nu}$ cannot be used
to make $\gamma_{\mu}=0$.
It can be shown that
$\delta^2$-terms are still absent in $H_{\mu\nu}$ and that the
equations (\ref{2.12})-(\ref{2.15}) with $\epsilon$ replaced by
$\eta$ are still valid. It can also be checked that
the tangential property
$S_{\mu\nu}\,n^{\nu}=\hat{G}_{\mu\nu}\,n^{\nu}=\hat{H}_{\mu\nu}\,n^{\nu}=0$
still applies.

As in the previous section
we introduce on $\cal {N}$ four intrinsic
coordinates $\{\xi^a\}$ with $a=1,2,3,4$ and the corresponding
holonomic tangent vectors $e_{(a)}=\partial /\partial \xi^a$. We
choose here  $e_{(1)}=n $ future-directed and therefore the other
vectors $e_{(A)}$ with $A=2,3,4$ are space--like. We then define
on $\cal{N}$ the basis $\{N, n, e_{(A)}\}$ where the transversal
is chosen light--like, perpendicular to the $e_{(A)}$'s and
oriented toward the future of $\cal{N}$. Thus $N$ satisfies
$N\cdot N=N\cdot e_{(A)}=0$ and $N\cdot n=\eta^{-1} =-1$.
The induced metric $g_{ab}=e_{(a)}\cdot e_{(b)}|_{\pm}$
on $\cal N$ is degenerate and reduces to
$g_{AB}=e_{(A)}\cdot e_{(B)}|_{\pm}$.

Of the
following two tensors,
$K_{ab}=-n_\mu\,e^\mu _{(a)|\lambda}\,e^\lambda _{(b)}$, 
${\cal K}_{ab}=-N_\mu\,e^\mu _{(a)|\lambda}\,e^\lambda _{(b)}$,
where the stroke has the same meaning as in section 1, 
only the second tensor describes transverse
properties and represents an extrinsic curvature. While $K_{ab}$
is purely intrinsic and therefore continuous across $\cal N$,
${\cal K}_{ab}$ is discontinuous across $\cal{N}$ with a jump
described by
$\gamma_{ab}\equiv\gamma_{\mu\nu}\,e^{\mu}_{(a)}\,e^{\nu}_{(b)}=
2[{\cal K}_{ab}] $. If we express the tensors $S^{\mu\nu}$,
$\hat{G}^{\mu\nu}$ and $\hat{H}^{\mu\nu}$ in terms of the tangent
basis $\{e_{(a)}\}$, and thus write $S^{\mu\nu} = S^{ab}
e^{\mu}_{(a)}e^{\nu}_{(b)}$, with similar expressions holding for
$\hat{G}^{\mu\nu}$ and $\hat{H}^{\mu\nu}$, then the equation for
$\hat{G}^{ab}$ is (see the eq.(31) of \cite{BI})
\be
2\, \hat{G}^{ab}=-(\gamma_{cd}\, g^{cd}_*)\,n^a n^b\,
-\gamma^{\dagger}\, g^{ab}_*\,
+ \{ g^{ac}_* \, n^b n^d\,+ g^{bc}_* \, n^a n^d \}\gamma_{cd}\, ,\n{3.5}
\ee
where one takes for $g^{ab}_*$ the matrix $g^{AB}$, inverse of $g_{AB}$,
bordered by zeros.
The equation for $\hat {H}^{ab}$ is derived from
the Gauss-Codazzi equations written on the basis $\{N, n,e_{(A)}\}$.
It leads to a complicated expression which contain the
intrinsic curvature tensor $R_{abcd}$, the intrinsic
three-dimensional curvature $K_{AB}$, and the extrinsic curvature
${\cal K}_{ab}$. It can be shown \cite{BI},\cite{BH} that a null
shell and an impulsive gravitational generally coexist with the
hypersurface $\cal {N}$ representing their space--time history.

As an illustration of this theory we consider a spherical null
shell propagating radially in the brane-cosmological model of
section 2. It is convenient to rewrite the line--element
(\ref{2.25}) in terms of the Eddington retarded or advanced time
coordinate $u$.
We introduce the constant sign
factor $\zeta$ which is $+1(-1)$ if the light cone ${\cal N}$ with
equation $u={\rm const}.$ is expanding (contracting) towards the
future.
The embedding requires that the function
$f_{k}(\chi)$ be the same on both sides of ${\cal N}$, but the
cosmological constant $\Lambda_5^{\pm}$ and the mass parameter
$m_{\pm}$
can differ. The components of
the normal $n$ and the transversal $N$ are
$n^{\mu}=\zeta\,\delta^{\mu}_r$ and $N^{\mu}=(1, -\zeta\,h(r)/2,
0, 0, 0)$, and the only non--vanishing components of ${\cal
K}_{ab}$ and $K_{ab}$ are ${\cal K}^{\chi}_{\chi} ={\cal
K}^{\theta}_{\theta} ={\cal K}^{\phi}_{\phi} = -\zeta\,h/2\,r$ and
$K^{\chi}_{\chi} =K^{\theta}_{\theta}=K^{\phi}_{\phi}=\zeta /r$.
Because $\cal N$ is a null cone, and therefore the light--like
signal is spherical--fronted, $\cal N$ cannot be the history of an
impulsive gravitational wave. In fact $\cal N$ is the history of a
spherical null shell which is simply characterized by its surface
energy density $\rho$ and surface pressure $P$.
In other words
the surface stress-energy tensor $S^{ab}$
has components $S^{11}=\rho$, $S^{AB}=P\,g^{AB}$ and $S^{A1} =0$
with
\be
\kappa_5\,\rho =\frac{3\,\zeta}{2\,r}\,[h(r)]\, -\frac{2\,\alpha\,
\zeta}{r^3}\,[h(r)]\,\left\{\frac{15}{2}\,\bar{h}(r)-\frac{(1-3k\,f_k
(\chi)^2)}{f_k(\chi)^2}\right\}\ ,\n{3.7}
\ee
\be
\kappa_5\,P=-\frac{9\,\alpha\,\zeta}{r^3}\,[h(r)]\,\left
(1+\frac{\bar{h}(r)}{2}\right )\ .\n{3.8}
\ee
When the
Gauss-Bonnet term is absent ($\alpha =0$) the surface pressure
vanishes ($P=0$) and the surface energy density of the null shell
is given simply by
\be
\kappa_5 \rho =-\frac{3\,\zeta}{2\,r^3}\,[m]\,
-\frac{\zeta\,r}{4}\,[\Lambda_5]\ .\n{3.9}
\ee
This is the analogue
in 5-D of the known expression for a similar null
shell propagating in the four-dimensional Schwarzshild-(anti)de
Sitter spacetime.

\end{document}